\def\year{2020}\relax
\documentclass[letterpaper]{article} 
\usepackage{aaai20}  
\usepackage{times}  
\usepackage{helvet} 
\usepackage{courier}  
\usepackage[hyphens]{url}  
\usepackage{graphicx} 
\usepackage{graphicx}  
\usepackage[linesnumbered, ruled, vlined]{algorithm2e}
\usepackage{amsmath}

\title{High Performance Depthwise and Pointwise Convolutions on Mobile Devices}
\author{
Pengfei Zhang, Eric Lo, Baotong Lu\\ \\
\Large{The Chinese University of Hong Kong}\\
\Large{\{pfzhang, ericlo, btlu\}@cse.cuhk.edu.hk}
}
 
\begin{document}

\maketitle

\begin{abstract}
Lightweight convolutional neural networks (e.g., MobileNets) 
are specifically designed 
to carry out inference directly on mobile devices.
Among the various lightweight models, 
depthwise convolution (DWConv) and pointwise convolution (PWConv) are their key operations.
In this paper,
we observe that the existing implementations of DWConv and PWConv are not well utilizing 
the ARM processors in the mobile devices,
and exhibit lots of cache misses under multi-core
and poor data reuse at register level.
We propose techniques to re-optimize the implementations of DWConv and PWConv based on ARM architecture.
Experimental results show that 
our implementation can respectively achieve 
a speedup of up to 5.5$\times$ 
and 2.1$\times$ against TVM \cite{tvm}
on DWConv and PWConv.
\end{abstract}

\noindent

\section{Introduction}

Recently, there is an increasing trend to carry out convolutional neural network (CNN) inference 
on mobile devices directly 
because of both privacy and real-time latency (user experience) requirements.
\cite{deepmon,mcdnn,mobilenetv1,mobilenetv2}.
However, 
since mobile devices are subjected to both computational and energy constraints,
recent research therefore puts effort on designing 
more lightweight ``mobile models'' 
that are composed of fewer layers 
and/or using less computational expensive operations.

In terms of CNN, 
examples of such lightweight mobile models include 
Xception \cite{xception}, MobileNetV1 \cite{mobilenetv1}, MobileNetV2 \cite{mobilenetv2}, MnasNet \cite{mnasnet}, EfficientNet \cite{efficientnet}, to name a few.

\begin{figure}
    \centering
    \includegraphics[width=0.6\columnwidth]{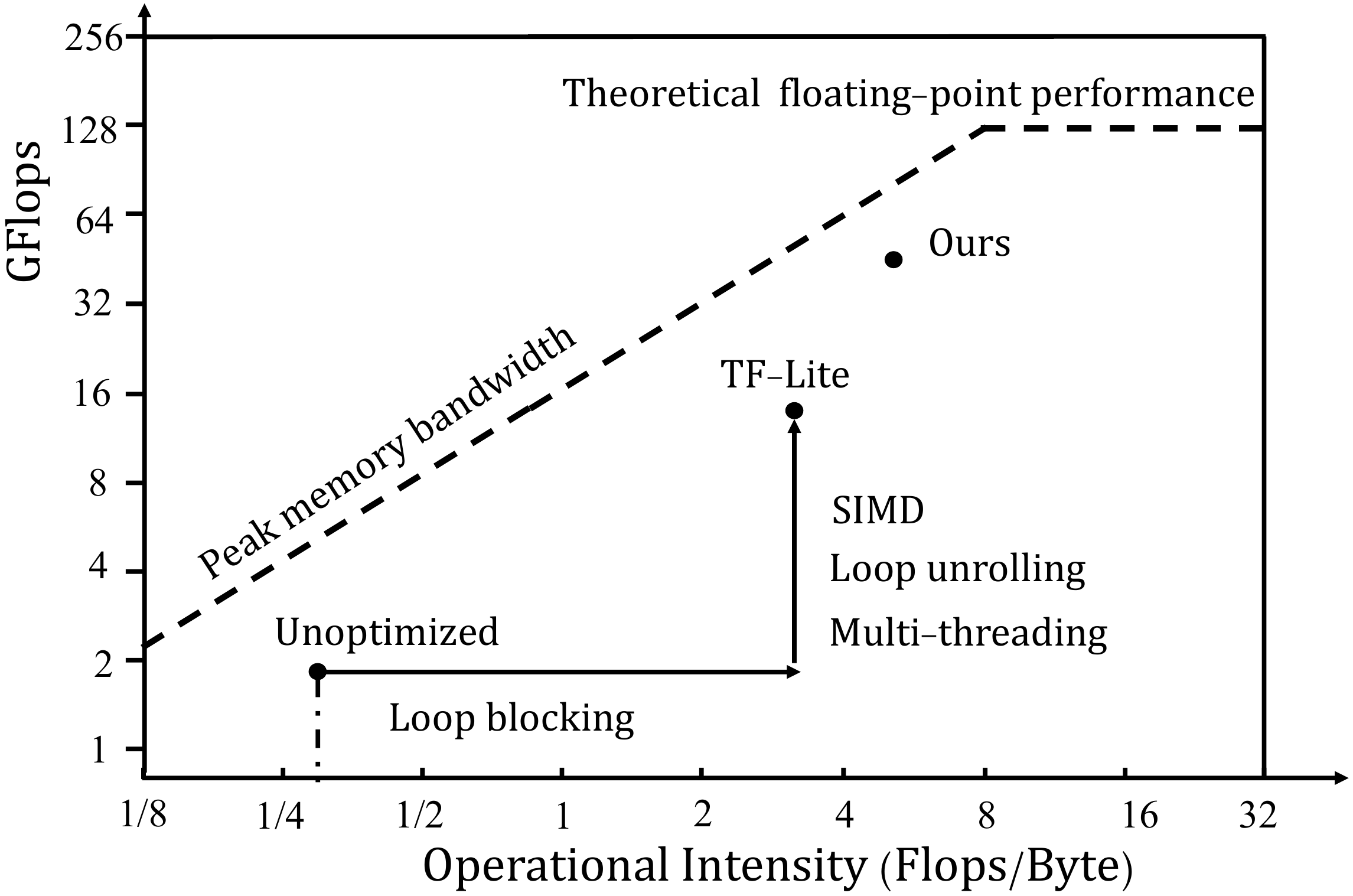}
    \caption{Roofline model for ARM Cortex-A57 with respect to MobileNetV1 inference}
    \label{fig:rooflinemodel}
\end{figure}


When optimizing the performance of a program with respect to a type of processors,
developers often use the \emph{roofline model} \cite{roofline} to guide their implementation.
Figure \ref{fig:rooflinemodel} shows the roofline model  of quad-core ARM Cortex-A57. The roofline (the dashed line) indicates the maximum achievable performance of any program under that processor.

Given the roofline model of a processor,
one can check whether her implementation 
has fully utilized that processor or not.
In Figure \ref{fig:rooflinemodel},
the point `Unoptimized' represents 
a naive C implementation of MobileNetV1 written by us.
The point `TF-Lite' represents 
the popular TensorFlow Lite 
binary 
compiled with math optimization, auto vectorization and linking to Eigen \cite{eigen} BLAS library.
Since TF-Lite is open source, 
it is known that it has already optimized using 
all the optimization tricks suggested in the roofline article (e.g., using SIMD intrinsics).
Unfortunately, 
even the popular TensorFlow Lite (TF-Lite)
is not fully utilizing the processor.
So, what is missing?

ARM processors get the lion's share of the mobile device processor industry \cite{armshare};
and DWConv and PWConv are the two most dominating operations in state-of-the-art mobile models
and they take up 90+\% of total inference time \cite{mobilenetv1,mobilenetv2,efficientnet}.
Therefore, the goal of this paper is to optimize 
\emph{depthwise convolution} (DWConv) and \emph{pointwise convolution} (PWConv) 
on ARM processors.
We observe there are two major issues that hurt the performance of DWConv and PWConv on ARM processors.

First, we point out that the {\bf existing DWConv and PWConv implementations are poor in core scalability},
which is against the trend of getting more cores 
in ARM processors (e.g., Huawei's latest mobile phone SoC chipset, Kirin 980, has eight ARM cores).
Second, we point out that the {\bf optimization tricks suggested in the roofline article are necessary but   insufficient for ARM processors}. 
Specifically, 
while both ARM and x86 processors can carry out 2 FMA (fused-multiply-add) instructions per cycle,
ARM processors 
can only load 1 register (from the cache) per cycle whereas x86 processors can load 4 registers per cycle.
In other words, 
while optimizing the cache miss and increasing parallelism 
could eliminate the major bottleneck on x86 processors,
on ARM processors those tricks could only shift the bottleneck to 
the traffic between the register and the cache.
Based on the above observations, 
we therefore develop high performance version of DWConv and PWConv for mobile devices.
Using techniques like \emph{loop rescheduling} \cite{loopscheduling}
and \emph{register tiling} \cite{registertiling},
our implementations are able to reduce 
the traffic between the cache and the memory 
\textbf{as well as}
the traffic between the register and the cache.
Experimental results show that 
our implementation can respectively achieve 
a speedup of up to 5.5$\times$ 
and 2.1$\times$ against TVM \cite{tvm}
on DWConv and PWConv,
which leads to a 46GFlops 
on ARM Cortex-A57 in terms of overall MobileNetV1 inference.

\section{Preliminaries} 
\label{sec_background}

ARM processors dominate the mobile device market.
Latest ARM processors all
support a 64-bit architecture, named ``AArch64".
AArch64 is a load-store architecture where data has to be loaded into the registers before
the operations take place.
AArch64 supports SIMD instruction and each core has 32 SIMD registers. 
Each SIMD register is 128-bit, which means each SIMD instruction can operate on 4 single precision numbers
simultaneously.
The predominate instruction used in model inference is the 
FMA (fused-multiply-add) SIMD instruction.
An FMA instruction requires 3 SIMD registers to fully operate.
Each FMA instruction carries out a 4-way SIMD multiplication, followed by a 4-way SIMD addition.

\subsection{Depthwise Convolution}
Depthwise convolution (DWConv) is a key operation in mobile models.
It takes three inputs: (i) a 3d array $\mathcal{I}$ (the input feature map) of size $H_i \times W_i \times C$, 
(ii) a 3d array ${\mathcal{F}}$ (the filter) of size $H_f \times W_f \times C$,
(iii) the stride $s$.
It produces a 3d array (the output feature map) $\mathcal{O}$ of size $H_o\times W_o \times C$.
In the above, $H$ and $W$ are the spatial height and width, $C$ is the number of channels.
The subscripts $i$, $f$ and $o$ refers to the input feature map, the filter, and the output feature map respectively.

\begin{algorithm}[h]
\caption{Unoptimized Depthwise Convolution}
\footnotesize
\label{algo:naive_dwconv}
	\KwIn{Input feature map $\mathcal{I}$, Filter ${\mathcal{F}}$, stride $s$;}
	\KwOut{Output feature map ${\mathcal{O}}$;}

	\For{$l = 0$ {\bf to} $H_o-1$}{
	\For{$k = 0$ {\bf to} $W_o-1$}{
	\For{$i = 0$ {\bf to} $C_i-1$}{
	\For{$n = 0$ {\bf to} $H_f-1$}{
	\For{$m = 0$ {\bf to} $W_f-1$}{
	${\mathcal{O}}_{l,k,i}$ += $\mathcal{I}_{l \times s+n, k\times s+m, i} \times {\mathcal{F}}_{n,m,i}$
	}
	}
	}
	}
	}
\normalsize					
\end{algorithm}

\begin{figure}
    \centering
    \includegraphics[width=0.6\columnwidth]{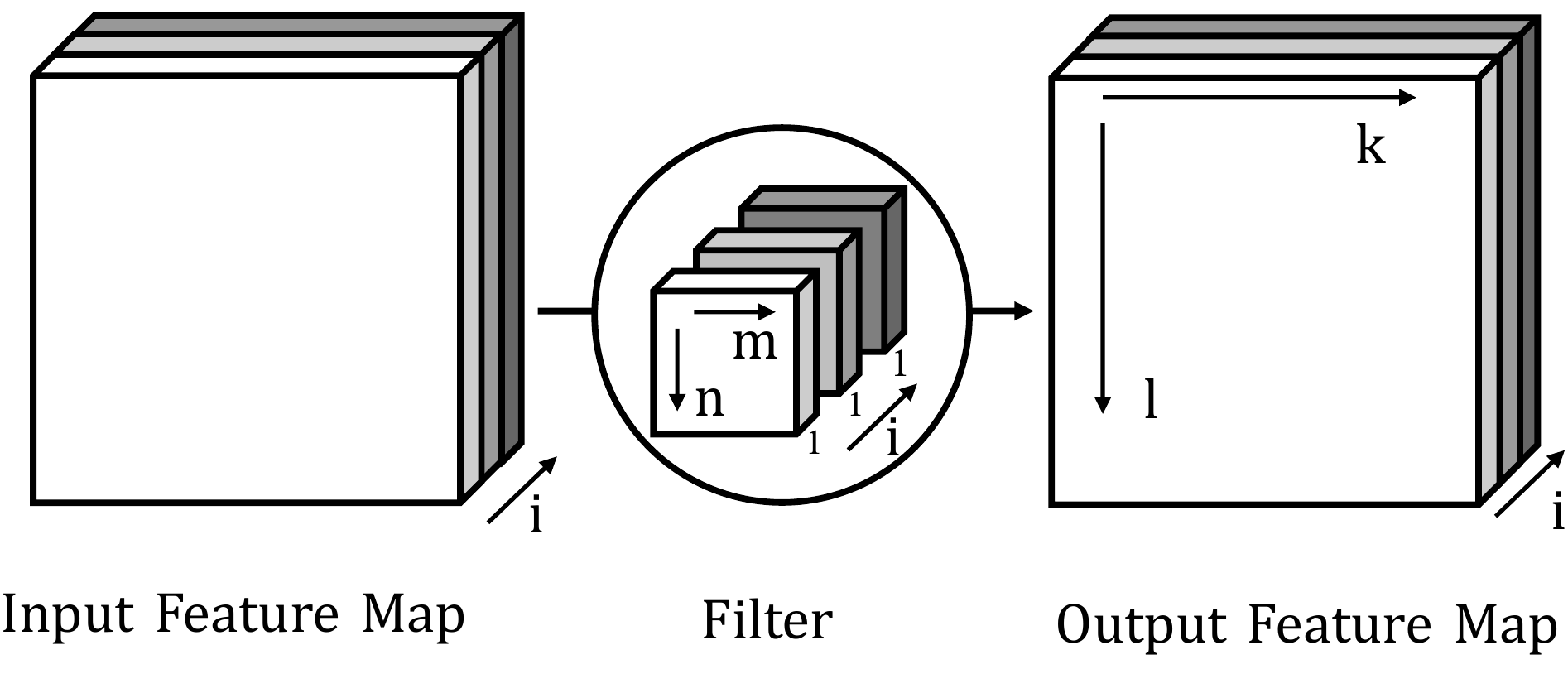}
    \caption{Depthwise convolution}
    \label{fig:depthwise}
\end{figure}

Figure \ref{fig:depthwise} illustrates the concept of depthwise convolution.
Algorithm \ref{algo:naive_dwconv} is its plain implementation,
which consists of 5 tightly-nested loops around a multiply-accumulate (MAC) statement (Line 6).
Referring to Figure \ref{fig:depthwise}, 
the implementation iteratively applies the filter (lines 4 and 5) per channel (Line 3),
and then repeats the task by moving the filter from left to right (Line 2) and then from top to bottom (Line 1).

\begin{algorithm}[h]
\footnotesize
\caption{Depthwise Convolution (TF-Lite)}
\label{algo:tflite_dwconv}
	\KwIn{Input feature map $\mathcal{I}$, Filter $\mathcal{F}$, stride $s$;}
	\KwOut{Output feature map $\mathcal{O}$;}

	\For{$l = 0$ {\bf to} $H_o-1$ {\bf in parallel}}{
	\For{$k' = 0$ {\bf to} $W_o/W_{o,b}-1$}{
	\For{$n = 0$ {\bf to} $H_f-1$}{
	\For{$kk = 0$ {\bf to} $W_{o,b}-1$}{
	\For{$m = 0$ {\bf to} $W_f-1$}{
    \For{$i' = 0$ {\bf to} $C / 4$} {  \tcp{Loop unrolling here.}
    $k = k' \times W_{o,b} + kk$ \\
    $V_I$=SIMD\_Load(
    $\mathcal{I}_{l\times s+n, k \times s + m ,i'\times 4\sim i'\times 4 + 3}$)\\
    $V_F$=SIMD\_Load(
    $\mathcal{F}_{n,m,i'\times 4 \sim i'\times 4 +3}$) \\
    $V_O$=SIMD\_Load(
    $\mathcal{O}_{l,k,i'\times 4 \sim i' \times 4 + 3}$)\\
    $V_O$=SIMD\_FMA($V_I$, $V_F$, $V_O$) \\
    SIMD\_Store($\mathcal{O}_{
    l,k,i'\times 4 \sim i' \times 4 + 3}$, $V_O$)
	}
	}
	}
	}
	}
	}
	
\normalsize					
\end{algorithm}

Algorithm \ref{algo:tflite_dwconv} shows the implementation of DWConv  in TF-Lite.
It mainly applies 4 tricks to improve its efficiency.
\begin{enumerate}
    \item {\sf Loop rescheduling and SIMD.} 
    Any permutation of the ordering (scheduling) of the loops would yield the same correct result
but with different efficiency.  
Furthermore, each channel of the filter can apply to corresponding channel of the input 
independently and thus \emph{in parallel}.
Consequently, Algorithm \ref{algo:tflite_dwconv} reschedules the innermost loop
to process the MAC across 4 channels using SIMD (lines 6--12).  

    \item {\sf Loop Unrolling.} The innermost loop actually possesses loop independence, meaning one iteration does not depend on its previous iteration.  In other words, the loop can be run in parallel.
    Consequently, the actual implementation of the innermost loop is \emph{unrolled} (or called as \emph{flattened}) \cite{loopunrolling}.  Loop unrolling not only improves ILP (instruction-level parallelism),
    but also reduces branch mis-prediction incurred by the test condition of each iteration.
    Algorithm \ref{algo:tflite_dwconv} however does not explicitly show the unrolled loop 
    for brevity.
    
    \item {\sf Loop Blocking.}  When involving matrix/tensor, loop blocking is often used to reduce cache misses \cite{loopblocking}. In TF-Lite, loop blocking is applied to the $k$ loop (Algorithm 1; Line 2)
    and it becomes the $k'$ loop in (Algorithm \ref{algo:tflite_dwconv}; Line 2)
    and the $kk$ loop in (Algorithm \ref{algo:tflite_dwconv}; Line 4).
    By doing so, the data loaded in the $k'$ loop (Algorithm \ref{algo:tflite_dwconv}; Line 2)
    could stay in the cache and get re-used again and again by the inner $n$ loop.
    
    \item {\sf Multi-threading.}  As real-time inference is getting more important,  TF-Lite also uses multiple cores to parallel the outermost loop (Line 1).  In other words, the blocks across the $l$ direction in Figure \ref{fig:depthwise} are generated by multiple cores.
    
\end{enumerate}

\subsection{Pointwise Convolution}

\begin{algorithm}[h]
\footnotesize
\caption{PWConv Implementation by MM}
\label{algo:tflite_pwconv}
	\KwIn{Input feature map $\mathcal{I}$, Filter $\mathcal{F}$;}
	\KwOut{Output feature map $\mathcal{O}$;}
    
    Mat A = $\mathcal{I}$.reshape([$G$, $C_i$]) \\
    Mat B = $\mathcal{F}$.reshape([$C_i$, $C_o$]) \\
    Mat D = $A \times B$ \\
    $\mathcal{O}$ = D.reshape([$H_o$, $W_o$, $C_o$])
\normalsize
\end{algorithm}

Another key component in mobile models is the pointwise convolution (PWConv). 
PWConv is a simple $1 \times 1$  convolution.
It takes as inputs: (1) a 3d input feature map $\mathcal{I}$ of size ($H_i \times W_i \times C_i$), and 
(2) a 4d filter $\mathcal{F}$ of size ($1 \times 1 \times C_i \times C_o$), 
and produces a 3d output feature map $\mathcal{O}$ of size ($H_o\times W_o \times C_o$), where $H_o = H_i$ and $W_o = W_i$.

Algorithm \ref{algo:tflite_pwconv} shows the implementation of PWConv in TF-Lite.
It essentially transforms the problem into 
a matrix-matrix (MM) multiplication problem $D=A\times B$,
where the 2d matrix $A$ is flatten from the 3d input 
$\mathcal{I}$, so that $A$ is a $G \times C_i$ matrix,
where $G= H_i \times W_i$ (Line 1); and
$B$ is a matrix of size $C_i \times C_o$ flatten from $\mathcal{F}$ (Line 2) since the first two dimensions are of size 1.

Since MM multiplication is a classic problem that has been well studied,
TF-Lite simply calls the high performance MM routine in a BLAS library \cite{blas}. 
MM multiplication implementations in BLAS are highly optimized with all the tricks (e.g., SIMD, loop rescheduling) mentioned above.
Recently, Google released an experimental matrix multiplication library named Ruy \cite{google_ruy}.  Ruy achieves good performance on small matrices (e.g., 100$\times$100) but its the performance on large matrices is poorer than BLAS. Since Ruy's code is still immature and flux,
we do not analyze it here but include that in our experiments.


\section{High Performance DWConv and PWConv} \label{sec:impl}

In this section,
we present techniques to optimize the implementations of DWConv and PWConv on ARM processors.
We will explain in detail why the existing ``well-optimized'' implementations 
are not efficient on ARM processors and propose our solutions.
One of the key elements there is about the notions 
of \emph{operational intensity} in the roofline model \cite{roofline}
and the notion of \emph{arithmetic intensity}  \cite{arithmeticIntensity}.


\begin{algorithm}
\footnotesize
\caption{\footnotesize{High Performance Depthwise Convolution}}
\label{algo:hp_dwconv}
	\KwIn{Input feature map $\mathcal{I}$, Filter $\mathcal{F}$, stride $s$;}
	\KwOut{Output feature map $\mathcal{O}$;}
	
	\SetKwFunction{kernel}{Kernel}
	\SetKwProg{Fn}{Function}{:}{}
	\For{$i' = 0$ {\bf to} $C/4 - 1$ {\bf in parallel}}{
	\For{$l' = 0$ {\bf to} $H_o/H_{o,b} - 1$}{
	\For{$k' = 0$ {\bf to} $W_o/W_{o,b} - 1$}{
		{\tt Kernel}($i'$, $l'$, $k'$, $s$)
	}
	}
	}
	\Fn{\kernel{$i'$, $l'$, $k'$, $s$}}{
	$\alpha = 0$ \\
	\If{$l'== 0$ \&\& $k' == 0$}{
    \For{$n=0$ {\bf to} $H_f-1$}{
    \For{$m=0$ {\bf to} $W_f-1$}{
 
        $V[\alpha]$ = SIMD\_Load($\mathcal{F}_{n,m,i'\times 4\sim i'\times 4+3}$)\\
        $\alpha$ += $1$
    }
    }
    }
    \Else{
    $\alpha = W_f \times H_f$
    }
    
    \For{$ll = 0$ {\bf to} $H_{o,b}-1$}{
    \For{$kk = 0$ {\bf to} $W_{o,b}-1$}{
    $l = l' \times H_{o,b} + ll$ \\
    $k = k' \times W_{o,b} + kk$ \\
    $V[\alpha]$ = SIMD\_Load($\mathcal{O}_{l,k,i'\times 4\sim i'\times 4+3}$) \\
    $\alpha$ += $1$
    }
    }
    
    \For{$ll = 0$ {\bf to} $H_{o,b}-1$}{
    \For{$kk = 0$ {\bf to} $W_{o,b}-1$}{
    $l = l' \times H_{o,b} + ll$\\
    $k = k' \times W_{o,b} + kk$\\
    \For{$n = 0$ {\bf to} $H_f-1$}{
    \For{$m = 0$ {\bf to} $W_f-1$}{
    \footnotesize{$V[\alpha]$ = SIMD\_Load(
    $\mathcal{I}_{l\times s+n, k\times s+m, i'\times 4\sim i'\times 4+3}$)} \\
    V[$H_f\times W_f + ll \times W_{o,b}+kk$] = SIMD\_FMA($V[\alpha]$,$V[n\times W_f + m]$, $V[H_f\times W_f + ll \times W_{o,b}+kk]$) \\
    }
    }
    }
    }
    
    $\alpha = W_f \times H_f$ \\
    \For{$ll = 0$ {\bf to} $H_{o,b}-1$}{
    \For{$kk = 0$ {\bf to} $W_{o,b}-1$}{
    $l = l' \times H_{o,b} + ll$ \\
    $k = k' \times W_{o,b} + kk$ \\
    SIMD\_Store($\mathcal{O}_{l,k,i'\times 4\sim i'\times 4+3}$, $V[\alpha]$) \\
    $\alpha$ += $1$
    }
    }
    }
\normalsize	
				
\end{algorithm}
\subsubsection{Roofline Model}
The roofline model \cite{roofline} is often used to understand the estimated performance of a given compute kernel 
running on a type of processor by showing the inherent hardware limitations, and potential benefit and priority of optimizations (e.g., locality, bandwidth, and different parallelization tricks).
The roofline model, however, focuses on cache misses.  In other words,
it focuses on the traffic between the cache and the memory
and assumes if the program is well optimized with little cache miss, the program could fully utilize the hardware.
The key metric inside the roofline model is ``operational intensity'' (OI),
which measures the average number of floating-point operations that can be carried out 
per byte of memory loaded from the memory.

\subsubsection{Arithmetic Intensity}
``Arithmetic Intensity'' (AI)  \cite{arithmeticIntensity}
measures the average number of floating-point operations that can be carried out 
per byte of memory loaded from the cache to the register.\footnote{We remark that there is a misconception online (e.g., Wikipedia) that OI is equivalent to AI.  
That misconception comes from the fact that cache miss is the major bottleneck on x86 processors and thus the traffic between the registers and the cache is immaterial after the cache bottleneck is resolved.  However, for ARM processors, it is not the case.}
This is exactly what we want to go after if the memory bottleneck can be removed.
Let $W$ be the number of arithmetic operations carried out, 
$\beta$ be the number of bytes transferred between cache and registers,
the arithmetic intensity $T$ is $\frac{W}{\beta}$.

Given a particular layer of convolution (e.g., DWConv), 
$W$ is a constant as it is dedicated by the problem definition and algorithm, 
a larger $T$ means the implementation is more efficient 
because there are fewer data transferred between the cache and the registers,
which implies the implementation is doing a good job in keeping the data in the register as long as it is necessary.

\subsection{Depthwise Convolution}

\subsubsection{Core Inscalability} 
Existing implementations of DWConv have poor scalability on the number of cores.  
Take TF-Lite implementation as an example (Algorithm \ref{algo:tflite_dwconv}), 
it picks the  $H_o$ dimension as the outer-most loop to apply thread parallelism (Line 1).
In other words, given $p$ cores, 
each core is assigned with a chunk of output feature map 
in size of $H_o/p \times W_o \times C$ to compute.

Since the chunk spans over all the output channels, each core has to copy the whole filter $\mathcal{F}$ of size $H_f \times W_f \times C$ into its tiny L1 cache.
In other words, 
when the input feature map, the filter, and the output feature map 
cannot all fit into the L1 cache,
the number of L1 cache misses will fly high.
Furthermore,
the situation exacerbates with the number of layers because 
the filters are getting larger when they appear deeper in the model.


\subsubsection{Poor AI} 
Although the implementation of DWConv in TF-Lite has good performance 
from the perspective of OI (and thus in terms of cache misses when we do {\bf not} use more cores),
its performance is next limited by its poor arithmetic intensity.
This is not an issue on x86 processors.
However, this is a big issue on ARM processor 
because ARM processors can only load 1 register per cycle 
while it can process 2 SIMD FMA instructions per cycle.
In other words, if we do not optimize the pipeline well,
the FMA instructions are always waiting for data to be loaded to the registers.

To be specific, we first analyze the AI of TF-Lite implementation (Algorithm \ref{algo:tflite_dwconv}).
Its inner-most loop is able to process 4 output elements in parallel by SIMD (Line 10).
In order to do so, however, 
it has to carry out 3 SIMD load instructions (Lines 7--9) 
to retrieve the filter, input and output respectively from cache to registers, 
and 1 SIMD store instruction to write back the updated output elements to L1 cache (Line 11). 
Thus, the arithmetic intensity of this implementation is $T_{tf}^{DW} = \frac{1\times 2\times 4~{\tt ops}}{4 \times 16~{\tt bytes}} = \frac{1}{8}$.
If the width $W_f$ of the filter
and the number of channels $C$ are small,
compilers may keep $W_f \times C$ elements of the filter
in the register for the $kk$ loop (Line 4).
To give TF-Lite such benefit of doubt,
we assume this happens 
and thus its arithmetic intensity can become  $T_{tf}^{DW} = \frac{1\times 2\times 4~{\tt ops}}{(3 + \frac{1}{W_{o,b}}) \times 16~{\tt bytes}} = \frac{1}{(3 + \frac{1}{W_{o,b}})\times 2} < \frac{1}{6}$.
Nonetheless, it is still a very poor number.



\subsubsection{Our implementation}

Algorithm \ref{algo:hp_dwconv} is our proposed implementation. 
To address the core inscalability problem, 
we re-schedule the loop order 
and picks the $C$ dimension as the outer-most loop to apply thread parallelism (Line 1).
This way, each core is assigned with a chunk of output feature map 
in size $H_o \times W_o \times \frac{C}{p}$ to compute.
Under such parallelism, since a chunk only spans $C/p$ output channels, each core needs to retrieve $H_f \times W_f \times C/p$ elements of the filter $\mathcal{F}$ to its L1 cache.
Compared with TF-Lite implementation that 
retrieves $H_f \times W_f \times C$ elements of the filter $\mathcal{F}$ to the L1 cache,
we fetch only $1/p$ of those in cache, which 
significantly reduce the cache misses and improve the core scalability.

To improve the arithmetic intensity,
we exploit different techniques to increase the reuse of the data in the register as much as we can.
The first technique we applied is  \emph{register tiling} \cite{registertiling}
(Lines 2 and 3).
It splits the filter $\mathcal{F}$ into tiles of size $H_f \times W_f \times 4$.
By doing so, a tile can be kept in the registers as long as possible. 
The kernel is used to compute the convolution results of a small output block of size $H_{o,b}\times W_{o,b} \times 4$.
$H_{o,b}$ and $W_{o,b}$ are set to ensure the output block stay in the registers across the {\tt Kernel}.
The kernel is skillfully tuned to increase its AI by reducing the traffic between the registers and the cache.
Specifically, lines 7 to 11 in the kernel aim to load the filter into the registers.
However, this load process is only done when $l'=0$ and $k'=0$ (Line 7),
meaning for the nested loops in lines 2 and 3, the filter is only loaded {\bf once} and stays in the registers for long.
Lines 14 to 19 in the kernel aim to load a specific output block of size $H_{o,b}\times W_{o,b} \times 4$ into the registers. 
Notice that this specific output block is only loaded {\bf once} 
and would never get re-loaded again.
Similarly, 
Lines 29 to 34 in the kernel aim to store the updated output block back to the cache.
Again this specific output block is only stored {\bf once},
as it would never get re-loaded for any further processing after it carries out the FMA in lines 20-27.

We now analyze the AI of our implementation.
That would help us to see why it outperforms the existing implementations.
It is easy to know the arithmetic operations are all inlined within the {\tt Kernel}.
In the {\tt Kernel}, the number of arithmetic FMA operations all lies in lines 18--25, 
which has 4 {\tt for} loops.
So, the FMA operation is carried out $\mathcal{W} = H_{o,b}\times W_{o,b} \times H_f \times W_f$ times.
Thus, the number of floating-point operations is $8 \times \mathcal{W}$, which will be the numerator in the AI.

The denominator of AI 
captures the number of bytes transferred between cache and registers.
For our implementation, it involves: 

\begin{enumerate}
    \item Loading the filter block once (Lines 7-11) across the nested two loops $l'$ and $k'$ (Lines 2 and 3) and reused $H_o/H_{o,b}\times W_o/W_{o,b}$ times. Thus, {\tt Kernel} incurs an average of $\frac{H_f \times W_f}{W_o/W_{o,b}\times H_o/H_{o,b}} \times 16$ bytes traffic between the registers and cache.
    
    \item Loading the output block once (Lines 14-19) and storing once (Lines 29-34) in the kernel.
    So, the traffic for output block in {\tt Kernel} is $H_{o,b}\times W_{o,b}\times 2\times16$ bytes. 
    
    \item Loading one SIMD register data of $\mathcal{I}$ in the inner-most loop (Lines 20-27).
    Thus, the traffic for $\mathcal{I}$ is  $16 \times H_{o,b}\times W_{o,b} \times H_f \times W_f = 16 \times \mathcal{W} $ bytes.
\end{enumerate}

Putting it all together, the AI of our implementation is:

\scriptsize
\begin{equation}
    T^{DW} = \frac{8 \cdot \mathcal{W}}
    {16(\frac{H_f \cdot W_f}{W_o/W_{o,b}\cdot H_o/H_{o,b}} + H_{o,b} \cdot W_{o,b} \cdot 2 + \mathcal{W})} 
    \label{eq:ai}
\end{equation}

\normalsize
\noindent

Since the size of the filter is either $3 \times 3$ or 
$5 \times 5$,
and the block sizes $H_{o,b}$ and $W_{o,b}$ are empirically set as 1 or 2
(they are set with the objective of saving some registers 
because we indeed apply loop unrolling to the 4 tightly-nested loops in Lines 18-25).
So, the term $\frac{H_f \times W_f}{W_o/W_{o,b}\times H_o/H_{o,b}}$ is negligible. 
Therefore, we rewrite equation (\ref{eq:ai}) as $T^{DW} = \frac{(H_f\times W_f)}{(2 + H_f \times W_f)\times 2} \ge \frac{9}{22}$,
which is obviously way larger than $T^{DW}_{tf}$.

\subsection{Pointwise Convolution Implementation}
\subsubsection{Core Inscalability}
TF-Lite's PWConv implementation by default calls the MM multiplication rountine in 
Eigen \cite{eigen}.
However, it is known that OpenBLAS \cite{openblas} has the best performance
and thus we set TF-Lite to use OpenBLAS instead.
Nonetheless, it is known that 
current matrix-multiplication implementations including OpenBLAS cannot scale well on multiple cores for deep learning workload \cite{zero_memory,deepcpu,mmscalability}.

\begin{algorithm}[h]
\footnotesize
\caption{Matrix Multiplication in BLAS Libraries}
\label{algo:blas_pwconv}
	\KwIn{Matrix A of size ($G\times C_i$), Matrix B of size ($C_i \times C_o$);}
	\KwOut{Matrix D of size ($G \times C_o$);}
	\For{$i' = 0$ {\bf to} $C_i/C_{i,b}$}{
	\For{$g' = 0$ {\bf to} $G/G_b$ {\bf in parallel}}{
	\For{$j' = 0$ {\bf to} $C_o/C_{o,b}$}{
	RTRA($i', g', j'$)
	}
	}
	}
\normalsize
\end{algorithm}

\subsubsection{Poor AI}
Algorithm \ref{algo:blas_pwconv} 
is the implementation of a BLAS MM routine (e.g., {\tt SGEMM} in OpenBLAS).
It has applied \emph{loop blocking} to increase data reuse in the memory hierarchy.
Its kernel 
is the function {\tt RTRA} (Line 4), which stands for \underline{R}egister \underline{T}iling \underline{R}euse block \underline{A}.
The logical view of {\tt RTRA} is depicted in Figure \ref{fig:tilingroutine} (left).
It first SIMD loads a block of matrix $A$, which is represented as \boxed{A}, into the registers (Line 2). 
\boxed{A} is of size $G_b \times C_{i,b}$.  
The elements of \boxed{A} stay in the registers across the \emph{j' loop} (Line 3 in Algorithm \ref{algo:blas_pwconv}) and are reused $C_o/C_{o,b}$ times.


Inside the function {\tt RTRA} (Figure \ref{fig:tilingroutine}),
Line 3 aims to stream a block of matrix \boxed{B} and a block of matrix \boxed{D} into the registers. 
\boxed{D} is of size $(G_b, C_{o,b})$ and 
\boxed{B} is of size $(C_{i,b}, C_{o,b})$.
A matrix multiplication between \boxed{A} and \boxed{B} is performed to update \boxed{D} (Line 4), and it costs $\frac{G_b \times C_{i,b} \times C_{o,b}}{4}$ FMA operations and the number of floating-point operations is $2\times G_b \times C_{i,b} \times C_{o,b}$.
Finally, the updated \boxed{D} has to be stored to the cache.

The AI of BLAS MM implementation is as follows.
The arithmetic operations are all inlined in the kernel {\tt RTRA}.
In routine {\tt RTRA}, its AI is:

\scriptsize
\begin{equation*}
    T^{PW}_{RTRA} = \frac{2 \times G_{b} \times C_{i,b} \times C_{o,b}~{\tt ops}}
    {(G_b \times C_{o,b}\times 2 + C_{i,b}\times C_{o,b} + \frac{G_b\times C_{i,b}}{C_o/C_{o,b}}) \times 4~{\bf bytes}} 
\end{equation*}

\normalsize
\noindent
Since AArch64 has 32 128-bit SIMD registers, in order to fully allocate the registers, $G_b$, $C_{i,b}$ and $C_{o,b}$ are usually set as $8$, $8$ and $4$ in the BLAS Libraries (e.g., OpenBLAS). 
Then, we can get $T_{RTRA}^{PW} = \frac{4}{3 + \frac{8}{C_o}}$. 
Note that the {\tt RTRA} kernel
has a poor AI because  \boxed{D} has to be transferred twice between the cache and the registers (one load and one store).

\begin{algorithm}[h]
\footnotesize
\caption{\footnotesize{High Performance Matrix Multiplication}}
\label{algo:hp_pwconv}
	\KwIn{Matrix A of size ($G\times C_i$), Matrix B of size ($C_i \times C_o$);}
	\KwOut{Matrix D of size ($G \times C_o$);}
    
    \For{$g' = 0$ {\bf to} $G/G_b$ {\bf in parallel}}{
    \For{$j' = 0$ {\bf to} $C_o/C_{o,b}$}{
    \For{$i' = 0$ {\bf to} $C_i/C_{i,b}$}{  
    RTRD($i', g', j'$)
    }
    }
    }
\normalsize					
\end{algorithm}

\subsubsection{Our Implementation}

\begin{figure} \centering
\includegraphics[width=1.0\columnwidth]{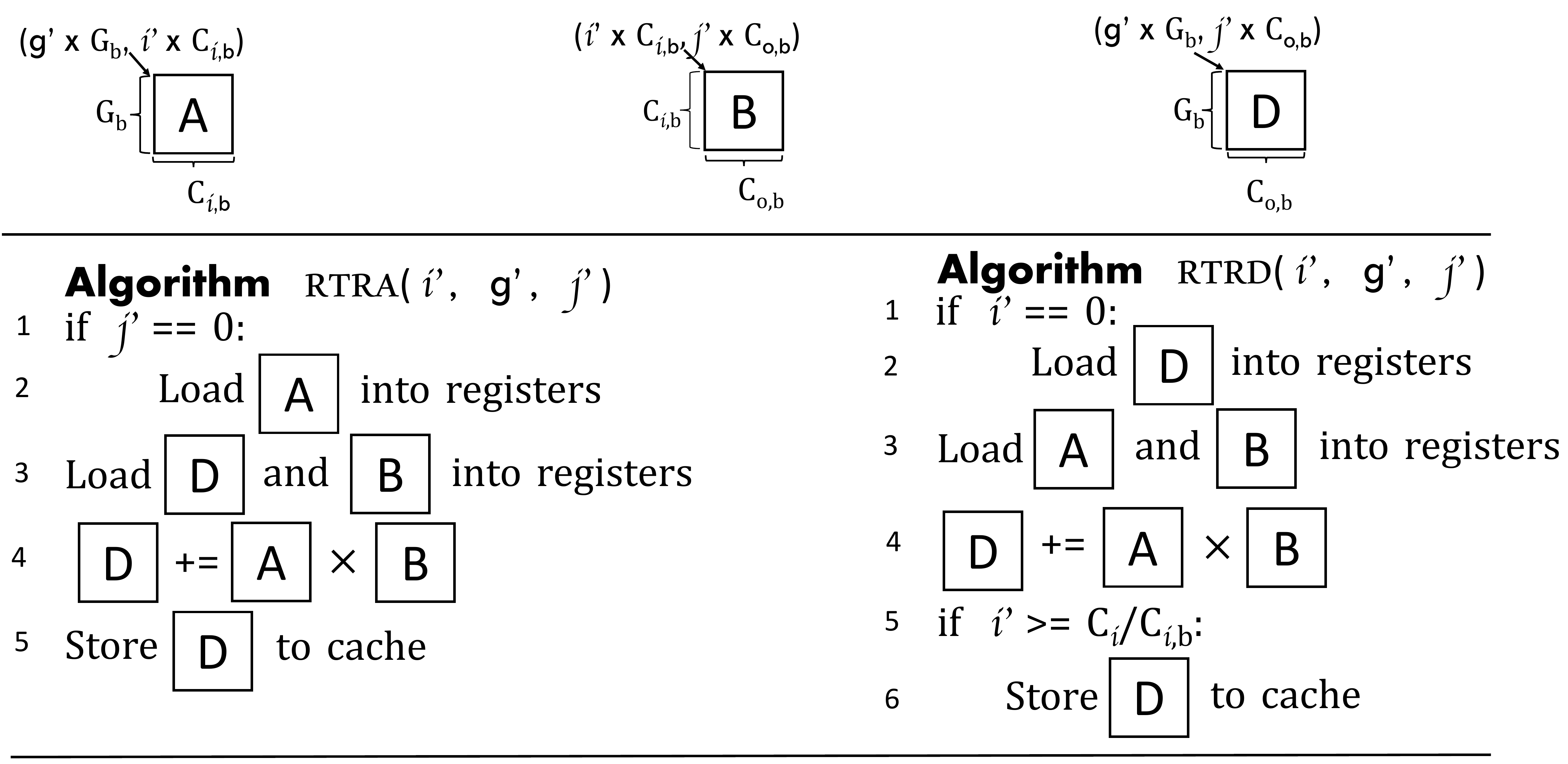}
\caption{{\tt RTRA} vs {\tt RTRD}}
\label{fig:tilingroutine}
\end{figure}

We propose another {loop blocking} method with better AI (Algorithm \ref{algo:hp_pwconv}).
It calls another kernel {\tt RTRD} (\underline{R}egister \underline{T}iling \underline{R}euse block \underline{D}), 
whose concept is listed in Figure \ref{fig:tilingroutine} (right). 
{\tt RTRD} first loads block \boxed{D} into the registers. 
The elements of \boxed{D} stay in the registers across the \emph{i' loop} (Line 3; Algorithm \ref{algo:hp_pwconv}) and are reused.
After that, it streams blocks \boxed{A} and \boxed{B} into the registers and then evaluates a small matrix multiplication to update \boxed{D} (Line 4).
Differ from {\tt RTRA}, {\tt RTRD} only stores the block \boxed{D} to the cache in the last iteration of loop $i'$.
Though this way is inefficient on x86 processor \cite{gotoblas}, 
it is very efficient for ARM processors because ARM processors is sensitive to AI.
The arithmetic intensity of {\tt RTRD} for MM multiplication is:

\scriptsize
\begin{equation*}
    T_{RTRD}^{PW} = \frac{2 \times G_{b} \times C_{i,b} \times C_{o,b}~{\tt ops}}
    {(G_b\times C_{i,b} + C_{i,b}\times C_{o,b} + \frac{G_b \times C_{o,b} \times 2}{C_i/C_{i,b}}) \times 4~{\bf bytes}} 
\end{equation*}
\normalsize

\normalsize

To fully allocate the registers, we can set $G_b = 8$, $C_{o,b}=8$ and $C_{i,b}=4$. 
Thus, $T_{RTRD}^{PW}=\frac{2}{1+\frac{8}{C_i}}$ 
it is about 1.5$\times$ larger than $T_{RTRA}^{PW}$, since $C_o$ and $C_i$ are often much larger than 8.
Of course, our actual implementation also includes all
the optimization tricks such as software prefetching, loop rolling etc.  
But we do not repeat them here.

\begin{figure*}
\begin{center}
\includegraphics[width=1.9\columnwidth]{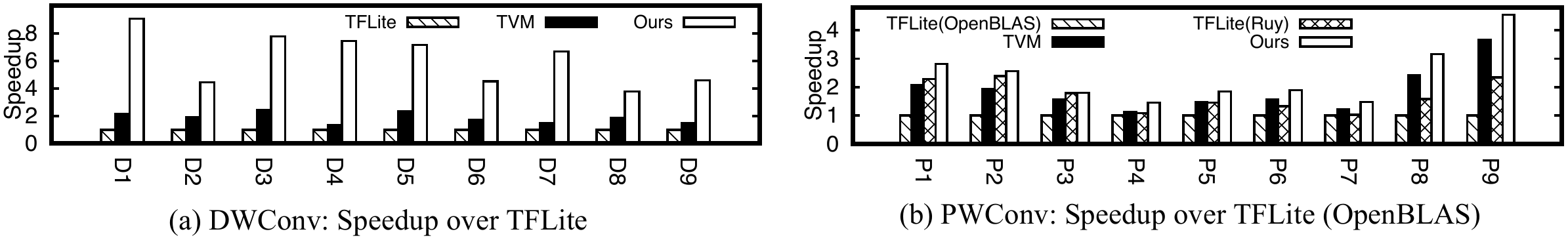}
\caption{MobileNetV1 (4/4 cores/threads)} \label{fig:mnv1}
\end{center}
\end{figure*}

\begin{figure*}
\begin{center}
\includegraphics[width=1.9\columnwidth]{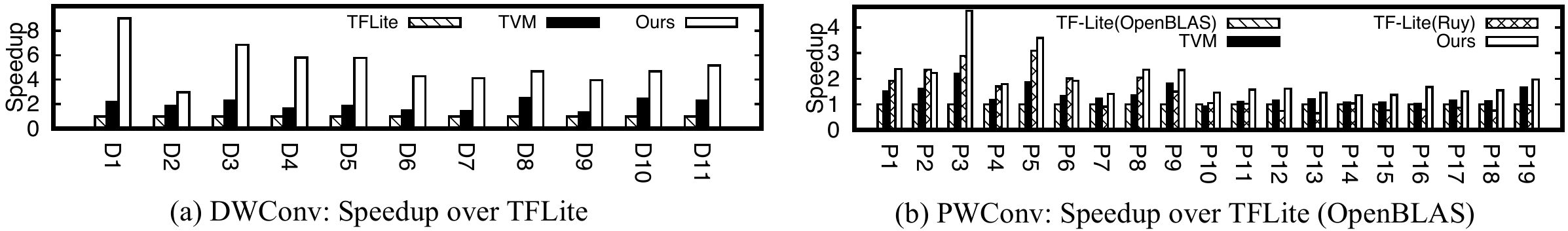}
\caption{MobileNetV2 (4/4 cores/threads)} \label{fig:mnv2}
\end{center}
\end{figure*}

\begin{figure*}
\begin{center}
\includegraphics[width=1.9\columnwidth]{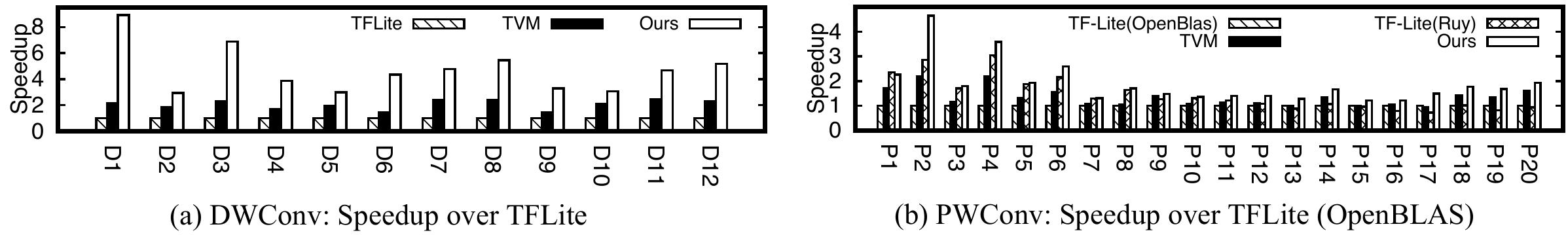}
\caption{MnasNet-A1 (4/4 cores/threads)} \label{fig:mnasnet}
\end{center}
\end{figure*}

\begin{figure*}
\begin{center}
\includegraphics[width=1.9\columnwidth]{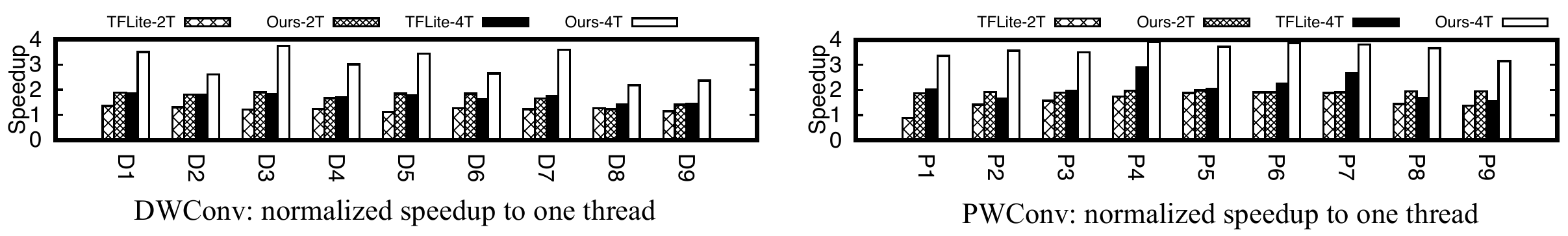}
\caption{Scaling behavior with increasing number of threads}\label{fig:scale}
\end{center}
\end{figure*}

\section{Experimental Evaluation}
In this section, we present performance results of our high performance depthwise and pointwise convolution on mobile devices.
We run our experiments on a 2.0GHz quad-core ARM Cortex-A57. 
Each core has 48KB L1 instruction cache and 32KB L1 data cache.
All cores share 2MB unified L2 cache.
We compare performance of our DWConv and PWConv implementations with two versions of TF-Lite, one links to OpenBLAS \cite{openblas} and the other one links to Ruy \cite{google_ruy}. 
In addition, we compare the performance with TVM \cite{tvm}. 
TVM implementations suppose to deliver performance as good as the performance offered by manually optimizing the implementation for a specific hardware.
The DWConvs and PWConvs operations in this study are extracted from
 MobileNetV1\cite{mobilenetv1}, MobileNetV2\cite{mobilenetv2} and  MnasNet \cite{mnasnet}. 
They are different in input size, output size and filter size.

\subsubsection{Performance}
Figure \ref{fig:mnv1} to
Figure \ref{fig:mnasnet}
show the speedup of our implementations (and TVM) with respect to TF-Lite, 
on different DWConv and PWConv extracted from MobileNetV1, MobileNetV2, and MnasNet-A1, respectively.
For example, in Figure \ref{fig:mnv1}, $D1$ to $D9$ refer to nine different DWConvs found in MobileNetV1.
Results show that our DWConv implementation outperforms TF-Lite at least by 2.9$\times$ and up to 9.0$\times$. 
In addition, our DWConv implementation outperforms TVM generated binaries
by at least 1.4$\times$ and up to 5.5$\times$,
showing that TVM is not able to reach the level of optimizations that we can achieve.


Our PWConv implementation achieves 1.3$\times$ to $5.1\times$ speedup over TF-Lite(OpenBLAS), 
which is essentially calling the OpenBLAS library for MM multiplication.
Our PWConv implementation also achieves up to 2.1$\times$ speedup over TF-Lite(Ruy), which uses the aggressively tuned library Ruy to implement PWConv. 
In addition, our PWConv implementation achieves 1.05$\times$ to 2.11$\times$ speedup over TVM,
which once again shows TVM is not able to reach the level of optimizations that we can achieve.

\subsubsection{Scalability}
In Figure \ref{fig:scale}, we compare the scalability of our DWConv and PWConv performances with respect to the number of cores.
We include TF-Lite (which uses OpenBLAS to implement PWConv) there for comparisons.\footnote{We do not include TVM here because TVM generates different binaries for different number of threads, making things incomparable.}
For space reasons, we only include the results from MobileNetV1
as results from MobileNetV2 and Mnasnet-A1 are largely similar.

From Figure \ref{fig:scale}, we see that our implementations scale better than TF-Lite.
We almost achieve perfect speedup when using 2 threads,
which is very promising because every parallel program has its serial part based on Amdahl's law.
When using 4 threads, the core instability of TF-Lite immediately manifest -- TF-Lite has only around 2$\times$ speedup on DWConv 
and 1.8$\times$ to 2.7$\times$ on PWConv.
In contrast, our implementations achieve 2.2$\times$ to 3.9$\times$ speedup on DWConv
and 3.2$\times$ to 3.9$\times$ speedup on PWConv.

\section{Related Work}

Most works 
on optimizing deep learning operations focus only on {conventional convolutions} 
\cite{zero_memory,mec,mkldnn,mmscalability} but not 
depthwise and pointwise convolutions appeared in mobile models.
To our best knowledge, 
this paper is the first to discuss the optimization of depthwise and pointwise convolutions on mobile processors.
In \cite{gfk}, there are treatments to  improve the performance of DWConv, but they focus on training and GPU,
whereas our focus is on inference and ARM.
TVM \cite{tvm} is a 
compiler stack for generating highly efficient binaries for deep network.
It supports CPU, GPU, ARM, and specialised accelerators.
Our experimental results show that binaries optimized by TVM not yet fully utilize the power of mobile processors.
BLAS libraries \cite{gotoblas,openblas,eigen} offer highly efficient implementations for PWConv.
However, we are able to show that they are still lacking on mobile devices.
%

%
%

\section{Conclusions and Future Work}

In this paper, we show that existing implementations of depthwise convolution
and pointwise convolution are not efficient enough on mobile devices.
The major reason is that those implementations 
have not considered the fact that ARM processors are getting more cores
as well as the latency gap between the load and FMA instructions in ARM processors.

To this end, we re-optimize the implementations of DWConv and PWConv specifically for ARM.
That is because ARM processors are dominating the mobile device market 
and there is an increasing demand to carry out inference directly on the mobile devices.
Experimental results show that 
our implementations can outperform industry-strength implementations from TF-Lite 
as well as optimized binaries generated from TVM.
Using MobileNetV1 as an example, our optimized implementation can carry out
inference at 46GFlops, a performance that is almost hitting the roofline of ARM processors.
The encouraging result also reveals one important future work for us.
Since TVM is a compiler framework for deep learning models,
our results indicate that we incorporate our techniques (e.g., register tiling) into TVM 
so to make it generate highly efficient binaries for mobile models on mobile devices.

\section{Acknowledgment}
This work is supported by Hong Kong General Research Fund (14200817,
15200715, 15204116), Hong Kong AoE/P-404/18, Innovation and Technology
Fund ITS/310/18.



\bibliography{reference} 
\bibliographystyle{aaai}

\end{document}